# Active deployable primary mirrors on CubeSat

Noah Schwartz[1], Maria Milanova[1], William Brzozowski[1], Stephen Todd[1], Zeshan Ali[1],
Lucie Buron[2], Jean-François-Sauvage[2,3], Charlotte Bond[1], Heather Bruce[1], Phil Rees[1],
Marc Ferrari[2], Donald MacLeod[1]

[1]UK Astronomy Technology Centre
Blackford Hill, Edinburgh EH9 3HJ, United Kingdom. Mail: noah.schwartz@stfc.ac.uk

[2]Laboratoroire d'Astrophysique de Marseille
Aix Marseille Univ, CNRS, CNES, LAM, Marseille, France;

[3]ONERA
29 avenue de la Division Leclerc, 92322 Châtillon, France;

**Abstract:** The volume available on small satellites restricts the size of optical apertures to a few centimetres, limiting the Ground-Sampling Distance (GSD) in the visible to typically 3 m at 500 km. We present in this paper the latest development of a laboratory demonstrator of a segmented deployable telescope that will triple the achievable ground resolution and improve photometric capability of CubeSat imagers. Each mirror segment is folded for launch and unfolds in space. We demonstrate through laboratory validation very high deployment repeatability of the mirrors <±5 µm. To enable diffraction-limited imaging, segments are controlled in piston, tip, and tilt. This is achieved by an initial coarse alignment of the mirrors followed by a fine phasing step. Finally, we investigate the impact of the thermal environment on high-order wavefront error and the conceptual design of a deployable secondary fitting inside 1U.

## INTRODUCTION

To fully exploit EO data, many commercial and scientific applications (e.g. Earth climate monitoring, natural disasters, defence and security) require very high-resolution images and as often as possible. However, combining both high spatial and temporal resolutions is, at the moment, out of reach at reasonable costs. Indeed, both requirements can only be achieved simultaneously by using multiple satellites in a LEO (Low Earth Orbit) constellation, which requires small individual satellites to lower costs. However, using small platforms (e.g. CubeSat) constrains the size of optical apertures, limiting the achievable spatial resolution. For instance, a 10 cm diameter telescope (typical maximum aperture on a 6U CubeSat) provides only 3 m resolution images from a 500 km orbit in visible wavelength (500 nm) due to the diffraction limit. Developing an optical aperture greater than 10 cm on a CubeSat represents a major opto-mechanical challenge.

In this paper, we present an innovative design for packaging (i.e. fitting inside the CubeSat volume), deploying, and controlling the position of the primary mirror of a segment space telescope. Each mirror segments is adjusted in piston, tip, and tilt (PTT) to reach the diffraction-limit of the telescope. We report on the development of a laboratory experimental setup to validate the deployment and alignment of the primary mirror (M1) segments. We also report, on preliminary studies investigating the impact of the thermal environment on high-order wavefront errors (not correctable by PTT), and on the conceptual design of a deployable secondary mirror that fits into a 1U volume when stowed.

# METER RESOLUTION FROM DEPLOYABLE TELESCOPE

## Deployable CubeSat

UK ATC's deployable 6U CubeSat will provide a 1 m ground resolution from LEO with a 30 cm optical aperture that can be deployed and aligned in space (diffraction-limit being 0.8 m at 500 km for a 500 nm wavelength) [1, 2]. Figure 1 indicate the currently unexploited region in terms of mass and ground resolution addressed by this work. By using a constellation of satellites, taking advantage of the cost reduction, it is possible to lower the waiting time between observations of a given ground scene.

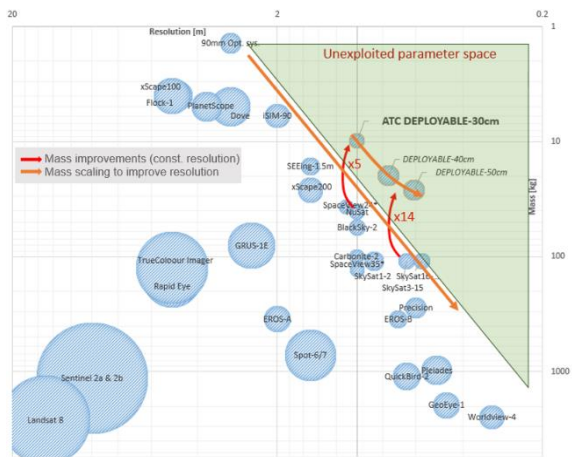

Figure 1: Image ground resolution (GSD) versus satellite mass and swath (bubble size) for deployable CubeSat and a selection of major satellites already in orbit.

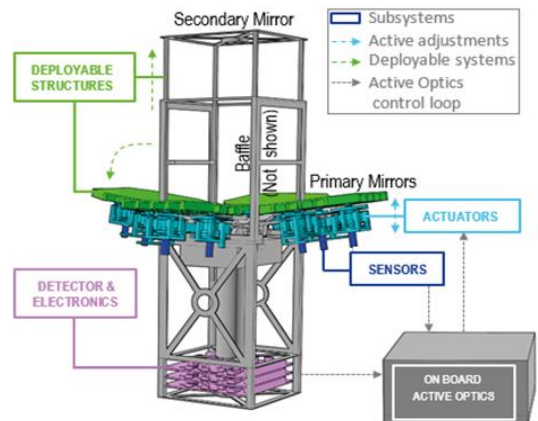

Figure 2: Simplified illustration of the payload concept: deployable structures (primary & secondary mirrors, baffle), actuators to adjust mirror positions, sensors to measure mirror positions, detector to assess image quality, and on-board computer to optimise mirror positions & image quality (i.e. active optics).

UK ATC's prototype provides a first demonstrator for a new payload technology: a standalone and automated deployable telescope dedicated to optical imaging. On a conceptual level (see Figure 2), deployable optical telescopes are composed of a primary mirror (M1, principal light-gathering surface), a secondary mirror (M2, folding light towards the detector), a baffle to stop unwanted light, and an active optics [AO] control to guarantee final optical quality. Finally, it often requires field correcting optics to compensate for field-dependent aberrations. Active and adaptive control of segmented primary mirrors a key enabling technology for space-based science (e.g. for instruments such as MIRI [3] on the James Webb Space Telescope) and for ground-based astronomy [4, 5]. Transposing this technology to the CubeSat market represents a major innovation.

## First error budget of a deployable CubeSat telescope

The mechanical repeatability of deployable structures is typically of the order of a few microns [1, 6]. To reach diffraction-limited performance (defined as λ/14) in the visible (500 nm), the total aberration level needs to be below approx. ≤35 nm. We incorporate to our system an active correction to compensate for static and slowly-varying aberrations. Actuators (Figure 2) provide movement to the optical surfaces in piston, tip, and tilt. Their extension must be greater than the deployment repeatability (i.e. several microns), and their resolution smaller than the allowable errors (i.e. tens of nm or less) [2].

On top of the M1 and M2 AO phasing residuals, error sources may also degrade the final resolution. These errors include the intrinsic optical quality (manufacturing, polishing errors), alignment (during MAIT, launch vibrations, thermal deformations), segment-related residual errors (differential piston, tip, and tilt, scalloping). Following [7], we set the requirement for each of the independent individual aberration contributors to ≤15 nm. Further system-level analysis would be required to define correlations between these error terms and refine their relative weight.

In order to reach this level of performance, we split the phasing process into 3 main steps. The first step consists in the initial deployment of the telescope in orbit from a folded and stowed configuration to a configuration in which the mirror segments are fully deployed. After deployment, the image (e.g. a PSF) should fall within the detector field of view, or better than 15 mrad for a 1° field of view detector. This precision is entirely achievable as demonstrated by the HighRes prototype [1, 2, 8]. The second step consists in a coarse phasing of the segments, orientating them towards the same light of sight. Several method have been investigate successfully in laboratory and simulations: using capacitive sensors or alternatively a blog detection algorithm. The final step consists in the fine phasing of the segments, where segments are co-phased to create a diffraction-limited optical system. For fine phasing results using image sharpness and phase diversity see [7, 8].

## M1 PROTOTYPE DESIGN AND LABORATORY EXPERIMENTATION

The prototype presented in this paper focuses on the main challenges: packaging, deploying, and aligning M1. For the purpose of the demonstrator, we limited the payload volume for M1 to 1.5U, other elements such as the secondary mirror, baffle, detector, etc. taking up the remaining of the 4U volume. A modular integration of the CubeSat systems was developed for the prototype laboratory demonstrator. This allowed easy access to all the various components for inspection, alignment and modification during the test programme. All structures are predominantly aluminium (including the mirrors) to ensure any differential thermal contraction effects are minimised.

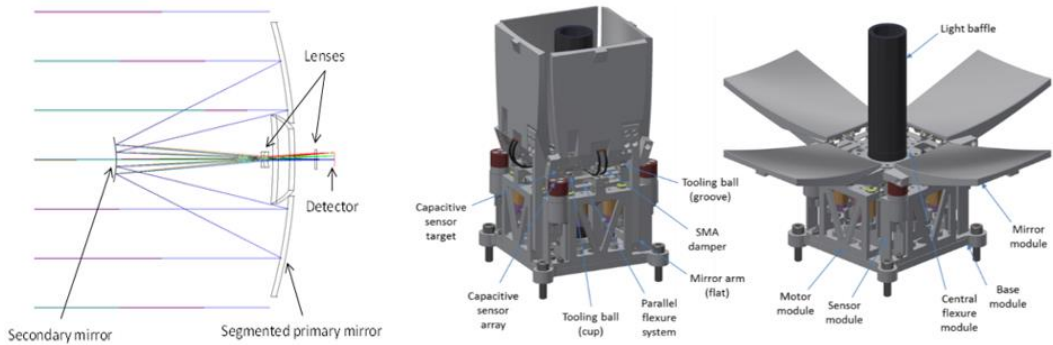

Figure 3: (Left) The optical design. Cassegrain telescope with a lens assembly to set the correct focal length and to provide aberration correction across the field-of-view (FoV). (Right) Overview of the mechanical design of the prototype CubeSat. Left: folded form for launch, fitting into a 1.5U volume. Right: deployed state reached after launch, deploying a 300 mm optical aperture.

The deployment motion is provided by an extension spring. It is mounted on a short lever arm that is designed to produce zero torque on the deployment axis when the mirror is in its final position. This is important so that the spring does not interfere with the tip-tilt adjustments on the mirror. Spring deployment mechanisms can cause undesirable shock or vibration issues when released. These undamped motions can cause repeatability issues in precision mechanisms. To prevent this from happening the mirror motion is controlled by the use of a case Nitinol wire [9] Shape Memory Alloy (SMA).

Three motors enable each mirror to be further manipulated in tip, tilt and piston. The motor chosen for this application is the Newport 8354 Tiny Picomotor. They have a 30 nm positioning resolution, a high actuation force (13 N), and have the ability to hold position while powered off.

## PRIMIRAY MIRROR DEPLOYMENT AND WAVEFRONT CONTROL

### Initial deployment

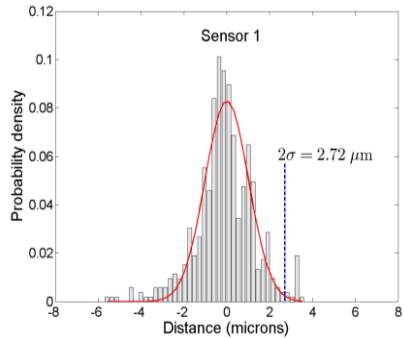

Figure 4: Deployment repeatability histograms of a sensor.

In order to produce a high optical quality, the mirrors need to be deployed with adequate repeatability and precision. To test the deployment performance, the mirrors were retracted and deployed into their kinematic mounts 500 times. Each time the mirror settled into its mount and a reading from 3 commercial off-the-self high-precision MicroEpsilon CapaNCDT CSE05 capacitive sensors was taken. Figure 4 shows the distribution of the measurement data for each sensor. The red line shows a normal distribution fit to the data with mean subtracted. From the figure it can be seen that in all instances the 2σ variation (95% confidence) around the mean is well within our required repeatability tolerance.

### Coarse phasing: close-loop control

For the coarse alignment, we use ZED-CAP [10] capacitive sensors in conjunction with the PicoMotor actuators to drive the mirrors into the desired position. We implemented an integrator control law: $u_{n+1} = u_n - gM_{control}s_k$, where $M_{control}$ is the control matrix, $g$ is the uniform scalar gain set to a small value, $u_n$ is the PicoMotors control vector at iteration step $n$, and $s_k$ is the capacitive sensor measurements. $M_{control}$ is calculated from the generalised pseudo-inverse of the interaction matrix which itself is measured by recording the response on the position sensors of each motor sequentially. Figure 5 shows the 3 capacitive channel measurements taken from one of the four mirrors.

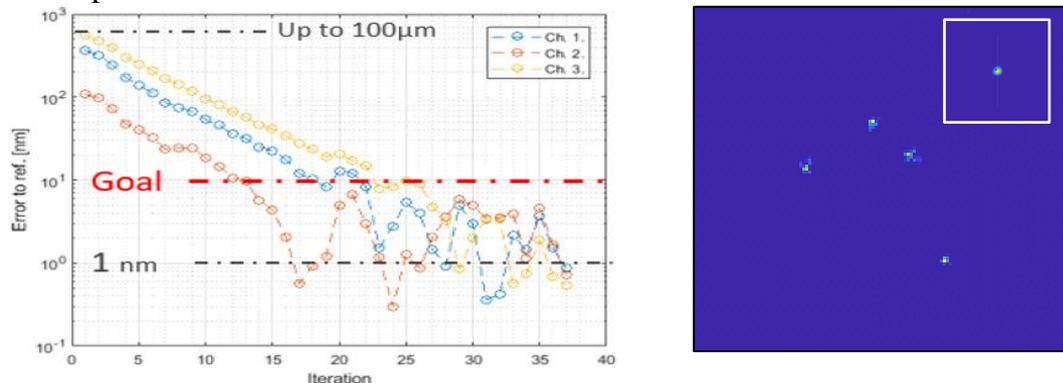

Figure 5: (Left) error as measured from the 3 displacement sensor channels as a function of iteration step. (Right) 4 individual PSFs before alignment; inset showing PSF aligned but not yet co-phased.

After an initial very fast convergence (here plotted with a low gain value to see convergence), the active optics loop appears to struggle to bring the error down below 10-20 nm, with noticeable oscillations. This has been identified to come mainly from the actuators used to drive the mirrors into position, and not the sensors themselves. In spite of the limited motor step size of approx. 30 nm, hysteresis and backlash (in addition to some limited amount of stiction in the system itself), we are able to converge to a final error <5 nm. This is below our requirement of 15 nm per mirror segment degree of freedom to reach diffraction-limited performance. It is important at this point to note that an initial calibration step is required to measure the absolute datum (i.e. reference mirror position for which perfect alignment is achieved). These positions are then used in the close loop operation as the reference.

## Coarse phase: blob detection

As an alternative strategy to using the on-board sensors we propose the use of a blob detection algorithm to perform the coarse phasing step. This method has the advantage of only relying on the detector images (i.e. where the final optical quality is required) without the need of a mechanical reference. In the blob detection algorithm, each petal is dithered individually using a small motor movement and a new image is taken (see Figure 6 and Figure 7). The difference between the new and previous image is computed. Finally, the PSF which belongs to the petal is identified and the petal is moved accordingly to align the PSF with the centre of the field of view. Preliminary simulation results show that tip-tilt (i.e. optical alignment) can be achieved with very high accuracy, with tip-tilt residuals of 50 nm RMS on average for bright stars. Five images are necessary to compute the differential images for the 4 petals, with a final image after alignment used for validation. Increased stability and robustness in terms of linear behaviour of the actuators or field distortions can be obtained by using small gain values, which would require more images. This leads to a trade-off between the number of image required and the final performance.

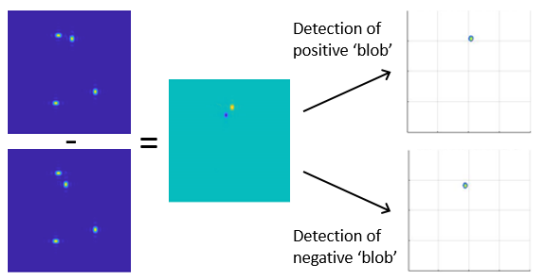

Figure 6: Principle of blob detection in differential image.

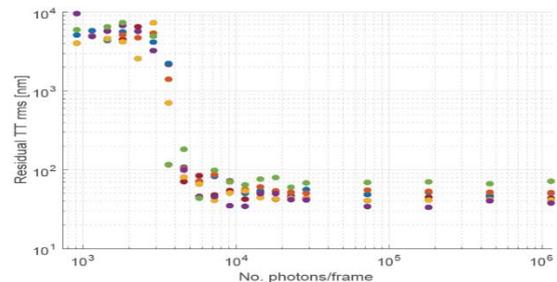

Figure 7: Tip-tilt residual after correction as a function of flux (RON=10 & single iteration alignment).

## Preliminary thermal analysis

Objective of this preliminary work is to study the thermal behaviour of the CubeSat in a representative LEO environment. The heat sources represented by Earth, Sun, and on-board elements create changing thermal gradients across the satellite structure and mirror petals. These gradients translate into mechanical deformations which in turn translate into optical aberrations. In particular, we want to evaluate two sources of deformation: the controllable deformation (i.e. Piston, Tip, and Tilt) to improve the AO control scheme, and high-order aberrations that are not in the controllable space. Our top-level requirement is to maintain the total high-order WFE (wavefront error) below 40 nm RMS. We conducted an analysis of the internal deformation of the mirror, for a normalised value of $\Delta T = 1K$ between the polished and rear face of the aluminium petal, using and end-to-end simulation tool (Nastran/Patran). This showed a curvature of 14 μm PV, twice the mechanical deformation. This deformation is well understood by considering a uniform plate deformation and the relationship: Sag = CTE x ThermalGradient x $L^2$ / (2 x width).

Figure 8 shows the extrapolated trend of the expected high-order optical aberrations for different materials and thermal gradients, using the linear relationship to CTE and ΔT. Keeping the high-order contribution below 40 nm RMS (approx. 100 nm PV) can be reached by decreasing the thermal gradient (using MLI and baffles) and with low CTE materials. The ceramic-based materials (e.g. cordierite) are a promising solution up to ΔT = 10K. The next steps of this analysis is to build a thermal model of the satellite (incl. main structure, secondary mirror, baffle & solar panel) to estimate thermal gradients and the time-evolution in a realistic environment. Finally, we will couple this analysis to simulations of a realistic active optics control to show the final correction capacity.

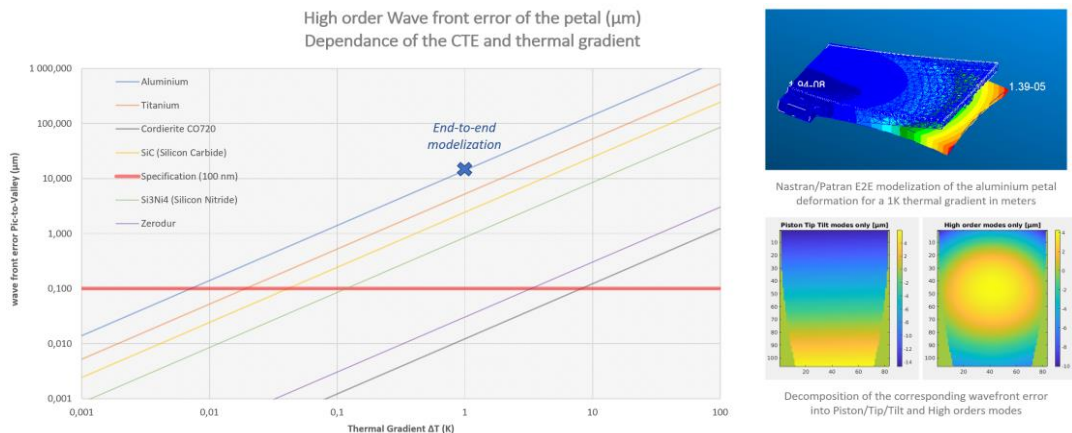

Figure 8: (Left) High order wavefront error of the petal; dependence of the CTE and thermal. (Right top) Example of Nastran/Patran E2E modeling for 1 mirror segment deformation. (Right bottom) Wavefront error for Piston, Tip, & Tilt only and remaining high-orders.

## DEPLOYABLE SECONDARY MIRROR

### Optical sensitivity for deployable CubeSats

The analysis of the optical design presented in section 2 showed that the optical tolerances were very tight. This is also true for the position of the secondary mirror M2. In order to analyse the sensitivity of M2 to misalignments, we investigate the impact of Defocus (i.e. a movement along the optical Z-axis), Decentre (i.e. a movement along the X/Y plane), and Tilt (a rotation around the X/Y axis) on aberration levels. After modelling and analysis, we set the M2 optical diameter to 50 mm to avoid vignetting and excessive mass. Both M1 and M2 have non-spherical surfaces. Because of symmetry, a rotation around the Z-axis has not impact on image quality.

Figure 9 shows the relative change in RMS wavefront error (WFE), which is subtracted from the RMS WFE of the nominal model (negligible even when using a simple field corrector designed with only a couple lenses). It shows the impact of defocus on final optical aberrations for 3 different M1-M2 distances (300, 400, and 500 mm), assuming the use of a field corrector. Unsurprisingly, increasing the M1-M2 separation helps reducing the sensitivity to defocus (i.e. we can tolerate larger defocus values). The 3 horizontal lines, represent respectively the diffraction-limit (i.e. 36 nm), the requirement (12 nm), and a goal (9 nm). For a M1-M2 separation of 300 mm, a defocus of ≤0.7 μm would still enable diffraction-limited imaging. This appears below the achievable precision of a single-use deployment system and would in practice require a fine movement actuator or stage to reach this value.

A similar analysis was performed on M2 decentre and tip-tilt (data not shown). This again demonstrates the need for large M1-M2 distance to reduce sensitivity, and shows that a decentre below ≤10 μm and a tip-tilt of $\Delta rX = \Delta rY < \pm 0.28$ mrad would still enable the system to reach the diffraction-limit. It is possible to slightly relax the tolerances by having actuation control in tip/tilt on top of having control of decofus.

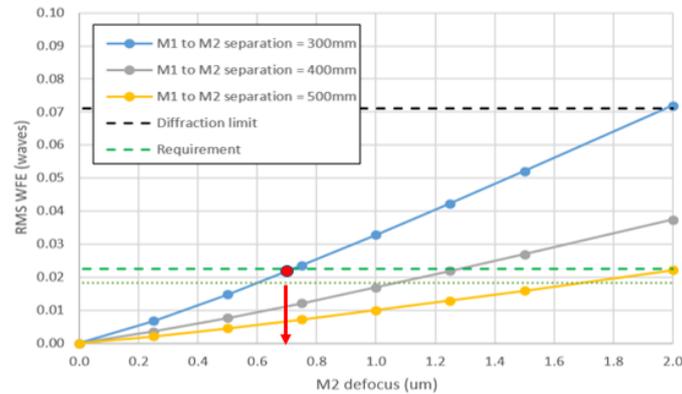

Figure 9: Impact of defocus on final optical aberrations for 3 different M1-M2 distances.

## Preliminary conceptual design

Nearly every design reviewed for this study utilises a reflective optical system and typically a variant of a Cassegrain layout (see for example [11, 12]). This design offers a large aperture and focal length by folding the light path within the CubeSat volume. In addition to deploying the primary mirror (M1) to increase the photometric power and angular resolution, there is a need to package and deploy the secondary mirror (M2) to minimize volume during launch and increase the focal length. As we have seen, increasing M1-M2 distance can have a dramatic effect on misalignment sensitivity.

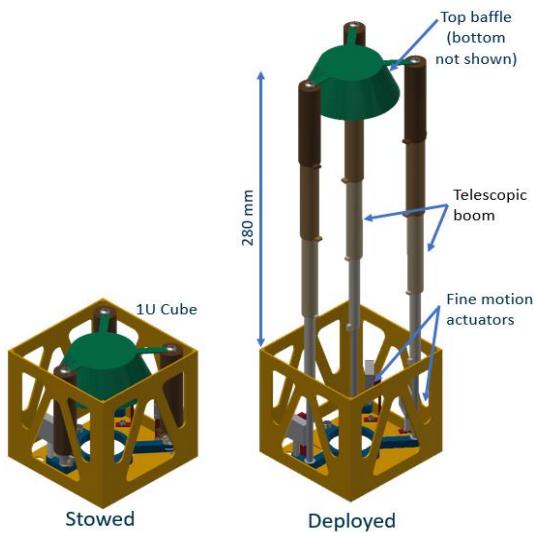

Figure 10: Deployable M2- preliminary conceptual design.

A trade-off study had shown telescopic booms to be most promising, see Figure 10. The deployment stages are attached to each other rigidly and semi-kinematically. They are secured with SMA for launch. The telescopic booms use preloaded springs and fine motion actuators for piston, tip/tilt control.

## CONCLUSION

In this paper we presented a concept for a high-resolution Earth Observation CubeSat using a deployable primary mirrors. A laboratory opto-mechanical prototype of this concept was developed and discussed in this paper. We demonstrated the feasibility of having a segmented primary mirror that can be deployed; its position measured and manipulated with the level of accuracy required for co-phasing in the visible. The primary mirror prototype was packaged within a 1.5U CubeSat volume. We validated that we can deploy the mirror segments with a repeatability of less than ±4.5 µm (typically less than ±3 µm). We demonstrated that the position of the mirrors, using the capacitive sensors and the linear piezo actuators, can be adjusted to within ≤10 nm of their optimal position; an accuracy sufficient to achieve diffraction-limited performance. Finally, we presented a sensitivity analysis of M2 misalignments, demonstrated that they are well within reach of current technology developed for phasing M1 segments, and proposed a conceptual design to deploy M2 using only a 1U volume.

The severe volume constrains - limited to 1.5U in our prototype version for fitting M1 - leads us to now consider a larger 4U alternative for the entire payload, which opens up new avenues in terms of opto-mechanical packaging, deployment strategies, and active optics correction. Using this larger volume, we now need to develop the concept further to reduce complexity (e.g. the number of active components), and take into account the volume for additional subsystems. Performance was also limited by the actuators used in the demonstrator, we are currently testing new COTS actuators. The conceptual design of M2 and baffle needs to be advanced further. Particular design emphasis will be placed on the deployment mechanisms and their accuracies, the resolution of the actuator mechanisms (required reach optimal optical quality), and future flight compatibility.